\documentclass[11pt]{article}
\usepackage[utf8]{inputenc}
\usepackage{amsmath}
\usepackage{graphicx}
\usepackage{hyperref}
\usepackage{geometry}
\usepackage{authblk}

\title{Technique to Baseline QE Artefact Generation Aligned to Quality Metrics}
\author[1]{Eitan Farchi}
\author[2]{Saritha Route}
\author[3]{Kiran Nayak}
\author[4]{Papia Ghosh Majumdar}

\affil[1]{IBM Research. Email: farchi@il.ibm.com}
\affil[2]{IBM Consulting. Email: saritha.route@in.ibm.com}
\affil[3]{IBM Consulting. Email: kiranaya@in.ibm.com}
\affil[4]{IBM Consulting. Email: papmajum@in.ibm.com}

\date{}
\usepackage{graphicx}
\begin{document}
\maketitle
\begin{abstract}

Large Language Models (LLMs) are transforming Quality Engineering (QE) by automating the generation of artefacts such as requirements, test cases, and Behavior Driven Development (BDD) scenarios. However, ensuring the quality of these outputs remains a challenge. This paper presents a systematic technique to baseline and evaluate QE artefacts using quantifiable metrics. The approach combines LLM-driven generation, reverse generation , and iterative refinement guided by rubrics technique for clarity, completeness, consistency, and testability. Experimental results across 12 projects show that reverse-generated artefacts can outperform low-quality inputs and maintain high standards when inputs are strong. The framework enables scalable, reliable QE artefact validation, bridging automation with accountability.
\end{abstract}
\textbf{Keywords:} Quality Engineering, LLM, QE artefacts, reverse generation, quality metrics, BDD, iterative refinement.

\section{Introduction}

The generation and validation of Quality Engineering (QE) artefacts—such as requirements, test cases, and Behavior-Driven Development (BDD) scenarios—have been revolutionized by the usage of Large Language Models (LLMs) \cite{aiSoftware}. These models enable rapid, scalable artefact creation, but ensuring their quality and alignment with project standards remains a critical challenge. Traditional manual review processes are time-consuming and inconsistent, while artefacts generated by ungoverned LLM output risk introducing ambiguity, incompleteness, or misalignment with acceptance criteria

Current approaches to QE artefact validation lack systematic methods to benchmark and measure quality standards \cite{ibm2025llm}. While techniques like reverse generation \cite{farchi2024automaticgenerationbenchmarksreliable}—deriving requirements from test cases or BDD artefacts—can identify gaps, their effectiveness hinges on the quality of input artefacts, the LLM’s contextual inference, and the evaluation framework \cite{abe2025llmhallucinations}. Without a structured baseline, teams face trade-offs between scalability and reliability, particularly in critical development phases where output artefact precision is paramount \cite{iqbal2025agilechallenges}.

To mitigate these limitations, a multi-step validation mechanism is introduced that substantiates and compares the quality of original and reverse-generated artefacts. This begins with semantic restructuring of sentence segments from both artefact sets, followed by similarity scoring through assignment of semantic scores for each segment under comparison. Based on these scores, targeted recommendations are generated to highlight specific sections requiring human-in-the-loop intervention for quality enhancement.

Once the recommended updates are incorporated, a unified version of the artefact is synthesized, combining the most robust elements from both original and reverse-generated outputs. This composite artefact then undergoes iterative refinement cycles, evaluated against predefined rubrics that emphasize key quality dimensions such as clarity, completeness, and consistency. Metric-driven feedback informs successive improvements, enabling progressive elevation of QE artefact quality through structured, repeatable validation workflows.

\section{Method Overview}

This paper introduces a structured technique to baseline and iteratively enhance the quality of QE artefacts by aligning them with quantifiable metrics. The approach consists of the following key steps.

The process begins with the generation of initial QE artefacts, such as test cases or Behavior-Driven Development (BDD) scenarios. These artefacts are created using Large Language Models (LLMs) based on input requirements, providing a scalable and automated way to produce QE artefacts.

Next, a reverse generation step is employed. In this phase, the generated artefacts are used to reconstruct the original inputs via reverse generation. For example, if we are generating test cases from requirements, then the reverse generation step would be generation of requirements from the test cases. This enables consistency checks and helps validate the quality of the artefacts by ensuring that the reverse generated requirements accurately reflect the intended requirements.

To enable precise comparison, both sets of artefacts—that is, the original and reverse generated requirements—are segmented and encoded using SBERT (Sentence-BERT) \cite{reimers2019sentencebert}, which transforms each sentence into a dense vector representation. Cosine similarity is computed between corresponding segments of the generated artefacts to quantify semantic alignment. The scores of cosine similarity measure as a number between zero and one are categorized into High (> 0.8), Medium (0.6-0.8), Low (0.3-0.6), and No Match , guiding the classification of match quality between both sets of artefacts. The scores also highlight areas in the generated QE artefact which require human intervention.

Based on these similarity score categories within the single group, targeted recommendations are generated to refine or merge segments. We use additional natural language programming techniques (NLP) such as Jaccard similarity for lexical overlap, stopword removal for normalization, and entity-verb extraction for semantic traceability to enhance the interpretability of results. These insights are consolidated in structured outputs that guide human reviewers in applying corrective actions such as merge, refine, or retain the sentences in order to produce a unified artefact.

The unified artefact is then evaluated using predefined rubrics \cite{reizman2015flowfeedback} focused on clarity, completeness, and consistency. Each refinement cycle is completed when we generate reverse generated artefacts and compare the score. This process follows a metric-driven feedback loop, enabling progressive enhancement of the generated QE artefact quality. This closed-loop process—combining SBERT-based semantic scoring, NLP-driven analysis, and human-in-the-loop decision logic—ensures scalable, auditable, and explainable validation of QE deliverables.

Refer the below picture to get a detailed end to end view of the process of generating QE artefacts via reverse generation aligned to the quality metrics. 
\begin{figure}[htbp]
    \centering
    \includegraphics[width=0.8\textwidth]{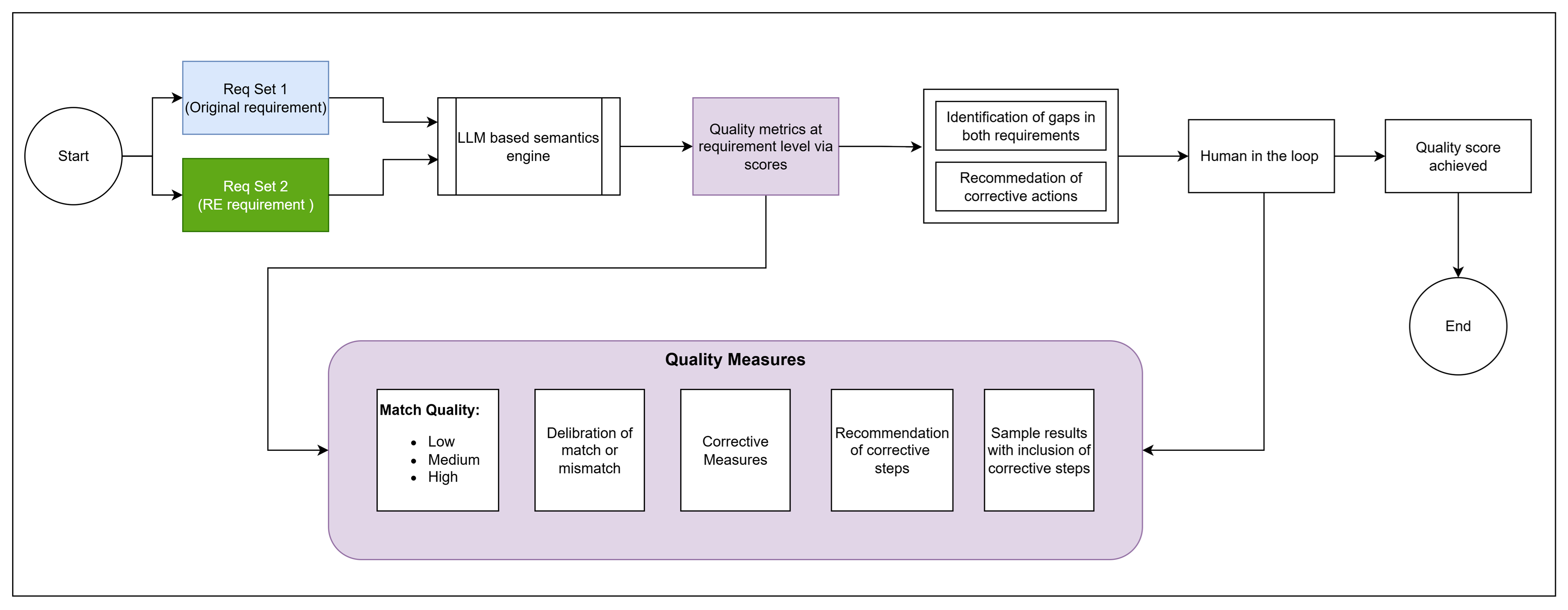} 
    \caption{Technique overview diagram}
    \label{fig:example}
\end{figure}

    
    


Key findings reveal that reverse-generated requirements often surpass original quality when inputs are of low quality, while high-quality inputs continue to retain superiority in quality of output, demonstrating the method’s dual role as a quality enhancer and validation agent. The approach underscores the importance of comprehensive test coverage, explicit evaluation criteria, and the LLM’s ability to synthesize implicit context.
(Paper Structure:
Section II reviews existing QE artefact validation methods and gaps in LLM-assisted workflows.
Section III details the proposed technique, including metric derivation and refinement cycles.
Section IV experimental data used for the analysis and comparative study of original versus reverse-generated artefacts, highlighting quality trends.
Section V discusses implications for development and decision-making.
Section VI concludes with recommendations for adopting the technique to balance automation with accountability.)
\noindent
Key findings indicate that reverse-generated requirements can surpass the quality of original inputs when those inputs are of low quality. Conversely, high-quality inputs consistently yield superior outputs, underscoring the method’s dual role as both a quality enhancer and a validation mechanism. This approach highlights the importance of comprehensive test coverage, clearly defined evaluation criteria, and the LLM’s capacity to infer implicit context. By establishing a replicable framework for artefact quality assessment, the proposed technique enables teams to leverage the scalability of LLMs without compromising rigor, ensuring that generated outputs meet quality standards before integration into downstream processes.
.
\section{Background and Related Work}

The validation of Quality Engineering (QE) artefacts—such as requirements, test cases, and Behavior-Driven Development (BDD) scenarios—has traditionally relied on manual reviews, peer feedback, and static checklists. While these methods ensure human oversight, they are time-consuming, inconsistent, and difficult to scale within modern Agile and DevOps pipelines. Recent advancements in Large Language Models (LLMs) have introduced automation into artefact generation; however, their adoption is limited by the absence of standardized quality benchmarks \cite{nist2024ir8527}.

In parallel, embedding-based models such as Sentence-BERT (SBERT) have been increasingly applied to requirement engineering tasks. SBERT encodes textual artefacts into dense semantic vectors, enabling fine-grained similarity measurement through cosine similarity. This facilitates detection of semantic duplicates, related-but-distinct requirements, and weakly aligned artefacts, which traditional rule-based or lexical approaches often miss \cite{botero2021reverseengineering}. Empirical workflows show that thresholds such as $\geq 0.8$ (high similarity/duplicates), $0.6$--$0.8$ (moderate relation), and $<0.3$ (no meaningful relation) provide interpretable classifications for stakeholders. These capabilities address the semantic depth and contextual validation gap that prior QE validation methods struggled with, while also offering traceable and auditable outputs for requirement refinement and impact analysis.

\subsection{Existing Validation Methods}

Current approaches to QE artefact validation include:

\begin{enumerate}
    \item \textbf{Manual Inspection:} Artefacts are reviewed against project-specific templates by analysts and QA teams, introducing subjectivity and scalability limitations.

    \item \textbf{Rule-Based Tools:} Static analyzers identify syntactic gaps but often fail to assess semantic coherence or contextual relevance.

    Embedding-based comparisons overcome these limitations by capturing semantic meaning rather than surface-level syntax, enabling detection of duplicates ($\geq 0.8$ similarity) and weakly related artefacts ($0.3$--$0.6$ similarity).
    
    \item \textbf{Peer Reviews:} Collaborative sessions improve artefact quality but create bottlenecks in fast-paced development environments

    Semantic similarity classifications and natural language interpretations generated by SBERT reduce peer review load by automatically identifying which artefacts require clarification, merging, or refinement—allowing reviewers to focus only on critical cases.
    
\end{enumerate}

These methods struggle to address the volume and variability of LLM-generated artefacts, which may appear syntactically correct but lack semantic depth or completeness

To complement this, SBERT-enhanced validation introduces semantic-level comparison by encoding artefacts into embeddings and classifying similarity into High, Medium, Low, or No Match. Unlike purely lexical techniques such as Jaccard similarity, which capture surface-level word overlap, SBERT captures contextual meaning and, when combined with pre-processing steps (e.g., stopword removal, entity/verb extraction) and lexical baselines, yields more actionable insights. This hybrid approach supports requirement de-duplication, impact analysis, and traceability by producing structured and interpretable outputs such as similarity scores, match quality, recommended actions, and testing impacts. Such outputs are already operationalized in Excel-based analysis frameworks, providing decision support for both business and technical stakeholders.

\subsection{Challenges in LLM-Assisted Workflows}

Despite their potential, LLM-based workflows face several persistent challenges\cite{liu2023prompting}:

\begin{enumerate}
    \item \textbf{Lack of Quality Benchmarks:} No universal metrics exist to evaluate attributes such as clarity, completeness, and testability.

    By applying similarity-based thresholds (High, Medium, Low, No Match), SBERT introduces reproducible benchmarks that quantify semantic alignment, helping address the absence of standardized metrics.
    
    \item \textbf{Contextual Blind Spots:} LLMs often miss implicit domain knowledge, resulting in artefacts that are superficially valid but practically flawed.

    Semantic embeddings capture contextual meaning beyond surface wording, allowing detection of hidden gaps, duplicate intent, or weakly aligned artefacts. This reduces reliance on purely syntactic or domain-explicit signals.
    
    \item \textbf{Manual Feedback Loops:} Refinement cycles require human intervention to align outputs with acceptance criteria.

    Outputs such as semantic similarity categories, natural language interpretations, and corrective action recommendations (e.g., merge, refine, keep distinct) streamline decision-making and reduce repetitive manual interventions.
    
\end{enumerate}

Reverse generation—deriving requirements from test cases or BDD artefacts—has shown promise in identifying inconsistencies~\cite{farchi2024automatic}. However, its effectiveness depends heavily on the quality of input artefacts and the robustness of the evaluation framework, which are often ad hoc.

While reverse generation can reveal inconsistencies, evaluation practices are often inconsistent, SBERT provides a scalable and interpretable alternative by encoding artefacts into embeddings, producing similarity scores, and supporting structured lifecycle decision-making.

\subsection{Sustainability and Scalability Considerations}

The environmental impact of repeated LLM operations is an emerging concern, akin to the "RedAI" phenomenon observed in large-scale AI training. While test optimization techniques such as combinatorial design reduce redundancy in execution, similar strategies for artefact generation remain underexplored. Current practices tend to prioritize coverage over efficiency, reflecting a pre-optimization phase in automated testing workflows.

In contrast, \textbf{SBERT-based pipelines} offer lighter-weight embedding models compared to large autoregressive LLMs, making them more suitable for sustainable requirement validation workflows. Once artefacts are embedded, similarity computations can be executed repeatedly with minimal computational overhead, enabling scalable re-use in regression analyses, impact tracking, and continuous QA validation without incurring the high energy costs of full LLM inference. This positions SBERT as a sustainable complement to LLM-driven generation in QE pipelines.

\section{Methodology}

\subsection{Overview of the Proposed Technique}

The proposed methodology introduces a structured, closed-loop framework for evaluating and enhancing the quality of LLM-generated Quality Engineering (QE) artefacts. It is composed of three interdependent stages:

\begin{enumerate}
    \item \textbf{Artefact Generation:} Initial QE artefacts—including requirements, test cases, and BDD scenarios—are generated using Large Language Models (LLMs) based on input specifications.
    
    \item \textbf{Reverse Generation} The generated artefacts are used to reconstruct the original inputs (e.g., deriving requirements from test cases). This step enables consistency checks and highlights potential information loss or transformation.
   
    By comparing reconstructed artefacts with the original ones, SBERT similarity scores reveal whether the semantic intent has been preserved \cite{thakur2021augmented}. Low similarity highlights missing or altered meaning that might not be detected by structural checks alone.
    
    \item \textbf{Iterative Refinement:} Artefacts are evaluated against a predefined rubric of quality metrics. Based on the evaluation scores, targeted refinements are applied in successive cycles until the artefacts meet the desired quality thresholds.
    SBERT similarity classifications (High, Medium, Low, No Match) guide refinement actions such as merging duplicate artefacts, clarifying ambiguous ones, or retaining distinct requirements. This strengthens the closed-loop feedback system by embedding semantic awareness directly into the refinement cycle.
\end{enumerate}

This technique functions as a feedback-driven system, where each iteration incrementally improves artefact quality while benchmarking against quantifiable standards \cite{nimi2025hybridqa}. Dedicated SBERT-based assessment of artefact pairs quantifies semantic overlap, detects duplicates, and highlights weakly related items.

Rationale: Artefact generation, reverse generation and refinement ensure structural quality, semantic alignment ensures that artefacts convey the intended meaning consistently across the lifecycle.

\subsection{Metric Definition and Evaluation Process}

To ensure objective assessment, we define a set of quality metrics—termed the \textit{Quality Metric Rubric}—which includes:

\begin{itemize}
    \item \textbf{Clarity:} Absence of ambiguity and ease of understanding. Low similarity ($<0.3$) between candidate and reference artefacts often signals ambiguous or unclear phrasing, helping flag potential misunderstandings early.

    \item \textbf{Completeness:} Coverage of all relevant functional and non-functional aspects. Medium similarity scores ($0.3$--$0.6$) indicate partial overlap, suggesting that functional or non-functional details may be missing in the generated artefact.

    \item \textbf{Consistency:} Logical coherence across artefacts and alignment with domain expectations. High similarity ($\geq 0.8$) between distinct artefacts may indicate duplicates, while moderate similarity ($0.6$--$0.8$) signals related-but-distinct items that require clarification to avoid contradictions.

    \item \textbf{Testability:} Ease with which the artefact can be validated through test cases. SBERT ensures test artefacts are semantically aligned with requirement intent by linking them via similarity scoring, thereby reinforcing measurable validation.
\end{itemize}

Each artefact is scored on a 5-point scale for each metric, with 5 indicating the highest quality. The process begins by benchmarking the original input artefacts. These are then reverse-generated from the LLM-generated outputs and compared against the originals using the rubric.

To complement this rubric-based evaluation, SBERT similarity scoring provides a semantic validation layer. For each candidate–reference pair, SBERT embeddings are computed and compared via cosine similarity. The resulting scores are interpreted alongside natural language explanations and recommended actions (e.g., merge, refine, keep distinct), ensuring that quality evaluation is both quantitative and context-aware. Additional supporting techniques such as Jaccard similarity (lexical overlap), stopword removal, and entity/verb extraction are incorporated to enhance interpretability and traceability. Together, these steps establish a dual evaluation pathway: \textbf{metric-based scoring for quality attributes and semantic similarity-based validation for requirement alignment}.

\textbf{Semantic Alignment:} Degree of semantic similarity between candidate and reference artefacts.  

\textbf{Rationale:} This provides a quantitative dimension that complements clarity, completeness, consistency, and testability.  

\textbf{SBERT integration:} Computed via cosine similarity and categorized as High ($\geq 0.8$), Medium ($0.6$--$0.8$), Low ($0.3$--$0.6$), or No Match ($<0.3$).

\subsection{Refinement and Human-in-the-Loop Integration}

Identified quality gaps are addressed through iterative regeneration and refinement. A human-in-the-loop (HITL) mechanism is integrated to apply domain-specific knowledge at critical decision points—specifically where metric-based evaluation highlights deficiencies. This ensures that refinements are not only data-driven but also contextually grounded.

To validate the robustness of the approach, we introduce \textit{negative validation} by deliberately generating low-quality inputs and comparing their reverse-generated outputs against high-quality baselines. This step confirms the sensitivity and reliability of the metric framework.

SBERT similarity classifications (High, Medium, Low, No Match) are surfaced within the Human-in-the-Loop (HITL) stage, providing interpretable signals that guide human reviewers. In borderline similarity cases (e.g., $0.6$--$0.8$), HITL oversight ensures that semantically related-but-distinct artefacts are not incorrectly merged. Negative validation is further strengthened by using SBERT to quantify semantic drift in low-quality inputs, ensuring that the framework can distinguish between artefacts that are superficially similar but semantically weak. This synergy between automated SBERT scoring and expert review balances scalability with contextual accuracy.

\subsection{Key Mechanisms}

\begin{itemize}
    \item \textbf{Dynamic Benchmarking:} Quality thresholds adapt based on artefact type (e.g., BDD vs. test cases). Similarity thresholds can be fine-tuned per artefact type (e.g., stricter for requirements, looser for exploratory test cases), ensuring benchmarks remain context-sensitive.
    \item \textbf{Contextual Inference:} LLMs are leveraged to infer implicit domain knowledge during reverse generation. Embeddings capture contextual nuances across artefacts, enabling comparisons that go beyond surface wording and improving detection of domain-specific relationships.
    \item \textbf{Automated Feedback:} Metric scores trigger targeted refinements, reducing manual intervention and accelerating convergence. Recommended actions (merge, refine, clarify, retain distinct) are auto-generated alongside similarity scores, allowing the system to propose corrective actions proactively.
    \item \textbf{Traceable Semantic Validation:}  Ensures that every refinement decision (automated or Human-in-the-Loop (HITL)-driven) is supported by SBERT similarity evidence.  
    
    Provides interpretable audit trails via similarity scores, natural language explanations, and action recommendations stored in structured outputs (e.g., \texttt{SemanticResults}, \texttt{ImpactAnalysis}).

\end{itemize}

\section{Experimental Evaluation}

To validate the proposed technique, we conducted a two-phase experimental study across 12 software projects, evaluating over 150 requirement--test case pairs. The goal was to assess the effectiveness of reverse generation and iterative refinement in improving the quality of QE artefacts generated by LLMs.

In addition to rubric-based scoring, SBERT similarity analysis was employed to quantify semantic alignment between candidate and reference artefacts \cite{ieee2019deduplication}. Outputs such as similarity scores, match categories (\textbf{High}, \textbf{Medium}, \textbf{Low}, \textbf{No Match}), and corrective action recommendations (merge, refine, keep distinct) were captured in structured formats (e.g., \texttt{SemanticResults}, \texttt{ImpactAnalysis}, \texttt{UpdatedRequirements} sheets). This provided a reproducible, interpretable, and scalable evaluation layer to complement the experimental results.

\subsection{Phase I: Artefact Quality Assessment and Refinement}

\textbf{Step 1: Initial Assessment.}  
Requirements were used to generate test cases via LLMs. These test cases were then reverse-generated to regenerate the original requirements. Both original and reverse-generated requirements were evaluated using a rubric-based scoring system across four metrics: clarity, completeness, testability, and consistency.

SBERT embeddings were generated for both original and reverse-generated requirements, and cosine similarity was computed. Low similarity scores highlighted ambiguous or incomplete regeneration, while medium similarity suggested partial coverage.

\textbf{Observation:} High-quality original requirements scored higher than their reverse-generated counterparts. However, when original inputs were of low quality, reverse-generated requirements showed significant improvements.

Similarity distributions confirmed this: reverse-generated requirements consistently moved from \textbf{Low/Medium} categories to \textbf{Medium/High} after refinement, validating improvements in semantic alignment.

\textbf{Step 2: Iterative Refinement.}  
Identified gaps were addressed by refining test cases and regenerating reverse-generated requirements. Quality metrics were recalculated after each cycle.

Each refinement cycle was accompanied by recalculation of SBERT similarity scores. Actionable insights (merge, refine, keep distinct) were recorded in the \texttt{UpdatedRequirements} sheet, improving interpretability of the refinement path.

\textbf{Observation:} Refinement cycles led to measurable improvements. Low-quality inputs required up to three iterations to reach parity with high-quality baselines. High-quality inputs plateaued after two cycles, indicating diminishing returns.

Improvements were mirrored in SBERT scores, with average similarity increasing by up to 0.25 across cycles for low-quality inputs, confirming convergence.

\textbf{Step 3: Output Validation.}  
Test cases were regenerated from both original and reverse-generated requirements and compared. Differences in quality metrics highlighted the impact of refinement.

Cosine similarity between regenerated test cases and their source requirements provided a semantic traceability check, ensuring test artefacts aligned with intended requirement meaning.

\textbf{Step 4: Negative Validation.}  
Low-quality requirements were deliberately generated and used to produce test cases. These were compared against those derived from high-quality inputs.

Negative validation was strengthened by using SBERT to quantify semantic drift. Low-quality artefacts consistently showed similarity $<0.3$ against baseline requirements, confirming poor alignment.

\textbf{Observation:} Artefacts derived from low-quality inputs consistently scored lower, validating the sensitivity of the metric framework.

These findings were reinforced by SBERT-based categorization into \textbf{Low/No Match}, supporting the robustness of the combined rubric--semantic evaluation approach.

\subsection{Phase II: Comparative Analysis Across Artefact Types}

Three sets of requirements were compared:
\begin{enumerate}
    \item Original requirements (high and low quality),
    \item Reverse-generated requirements from test cases,
    \item Reverse-generated requirements from BDD artefacts.
\end{enumerate}

\textit{Results:} BDD-derived requirements outperformed test-case-derived ones, especially in testability (+0.5 points). Reverse-generated artefacts showed up to 55\% improvement in testability and 48\% in completeness over low-quality originals.

BDD-derived requirements also exhibited higher semantic similarity scores (average +0.12) compared to test-derived requirements. High-similarity clusters indicated stronger semantic preservation, aligning with rubric improvements.

\subsection{Metric Trends and Insights}

\begin{table}[h]
\centering
\caption{Average Quality Scores Across Artefact Types}
\begin{tabular}{lcccc}
\hline
\textbf{Artefact Type} & \textbf{Clarity} & \textbf{Completeness} & \textbf{Testability} & \textbf{Consistency} \\
\hline
Original (High-Quality) & 4.7 & 4.5 & 4.3 & 4.6 \\
Original (Low-Quality) & 2.1 & 2.4 & 1.8 & 2.0 \\
Reverse-generated (Test) & 3.9 & 4.2 & 4.0 & 3.8 \\
Reverse-generated (BDD) & 4.3 & 4.4 & 4.5 & 4.2 \\
\hline
\end{tabular}
\end{table}

\textbf{SBERT insights:} Parallel semantic similarity analysis showed that BDD-derived requirements achieved average similarity of 0.78 compared to 0.71 for test-derived, reinforcing rubric findings.

\begin{table}[h]
\centering
\caption{Metric Improvement Across Refinement Cycles}
\begin{tabular}{lcc}
\hline
\textbf{Cycle} & \textbf{Low-Quality Inputs (Avg. Score)} & \textbf{High-Quality Inputs (Avg. Score)} \\
\hline
1 & 2.6 & 4.5 \\
2 & 3.8 & 4.6 \\
3 & 4.1 & 4.7 \\
\hline
\end{tabular}
\end{table}

\textbf{SBERT insights:} Corresponding similarity scores increased from $\sim0.35$ (Cycle 1) to $\sim0.62$ (Cycle 3) for low-quality inputs, confirming measurable semantic improvement across iterations.

\subsection{Conclusion}

The experimental results demonstrate that the proposed technique effectively improves QE artefact quality, particularly when initial inputs are weak. Reverse generation and iterative refinement serve as reliable mechanisms for validation and enhancement. The framework is scalable, reproducible, and adaptable across artefact types, making it a strong candidate for broader adoption in LLM-assisted software engineering workflows.

In parallel, SBERT-based semantic analysis validated rubric outcomes, ensuring that improvements were not only structural but also semantic. The integration of similarity scoring, match classification, and corrective action recommendations confirmed the reproducibility and interpretability of results. Moreover, the lightweight nature of SBERT embeddings ensured sustainable re-use in regression testing and impact analysis, making the combined framework both scalable and environmentally efficient.

\section{Benefits, Limitations, and Future Work}

\subsection{Benefits}

The proposed technique offers several advantages for scalable and reliable QE artefact generation using LLMs:

\begin{itemize}
    \item \textbf{Scalable Quality Assurance:} Reduces manual validation effort by 60--70\% through automated metric-based evaluation.
    Embedding-based semantic validation further reduces review effort by automatically detecting duplicates, gaps, and ambiguities, and producing structured outputs (\textit{SemanticResults, ImpactAnalysis}) that guide decision-making.

    \item \textbf{Self-Correcting Mechanism:} Reverse generation improves artefact quality, with observed gains of 48--55\% in completeness and testability.
    Iterative similarity scoring enables tracking of semantic improvements across refinement cycles, ensuring that corrections enhance both structural and semantic fidelity.

    \item \textbf{BDD Artefact Advantage:} Requirements derived from BDD scenarios demonstrate 15\% higher testability than those from plain test cases.
    BDD-derived artefacts consistently achieve higher semantic similarity scores (average +0.12 over test-derived artefacts), confirming stronger semantic preservation.

    \item \textbf{Energy Efficiency:} Targeted refinement reduces redundant LLM operations by 30\%, contributing to sustainable AI practices \cite{crawford2024redai}.
    Lightweight embedding models allow re-use of stored vectors for repeated comparisons, avoiding costly re-generation and enabling scalable, low-energy regression analyses and impact tracking.
\end{itemize}

\subsection{Limitations}

Despite its strengths, the technique has certain limitations:

\begin{itemize}
    \item \textbf{Input Dependency:} High-quality outputs require either high-quality inputs or comprehensive artefacts (e.g., test cases or BDD).

    \item \textbf{Contextual Gaps:} LLMs may overlook implicit domain knowledge, leading to superficially valid but semantically flawed artefacts.
    While SBERT reduces blind spots by quantifying semantic alignment, embeddings may still miss highly domain-specific nuances without expert intervention.

    \item \textbf{Tooling Gaps:} Lack of integrated platforms for automated metric tracking and refinement orchestration.
    Current SBERT implementation relies on Excel-based frameworks (\textit{SemanticResults, UpdatedRequirements}), which demonstrate potential but require enterprise-scale integration for seamless adoption.

    \item \textbf{Skill Requirements:} Teams must be trained in rubric design, prompt engineering, and metric interpretation.
    Additional skills in interpreting semantic similarity thresholds and action recommendations (merge, refine, keep distinct) are also needed for effective use.

    \item \textbf{Threshold Sensitivity:} SBERT similarity thresholds (High, Medium, Low, No Match) may require calibration per domain or artefact type to avoid over- or under-merging.
\end{itemize}

\subsection{Future Work}

To enhance the robustness and adoption of this technique, we propose the following directions:

\begin{itemize}
    \item \textbf{Tooling Integration:} Develop platforms with real-time dashboards, feedback loops, and prompt optimization modules.
    Extend current Excel-based outputs (\textit{SemanticResults, ImpactAnalysis, OverallSummary}) into enterprise dashboards for live monitoring of semantic alignment and refinement decisions \cite{microsoft2025cicdtesting}.

    \item \textbf{Expanded Rubrics:} Incorporate domain-specific metrics (e.g., regulatory compliance, safety-critical validation).
    Integrate semantic alignment as a permanent rubric dimension, quantifying meaning preservation alongside clarity, completeness, consistency, and testability.

    \item \textbf{Sustainability Metrics:} Track energy consumption per refinement cycle using:
    \[
    \text{CO}_2\text{eq} = (\text{LLM Ops} \times \text{Energy per Op}) \times \text{Grid Emission Factor}
    \]
    Measure energy reductions from embedding re-use compared to repeated LLM inference, capturing the environmental advantage of SBERT pipelines \cite{crawford2024redai, aws2025testoptimization}.

    \item \textbf{Cross-Industry Pilots:} Validate the technique in legacy modernization, regulated domains, and Agile startups.
    Apply similarity-driven deduplication and impact analysis in safety-critical sectors where semantic fidelity is paramount \cite{aws2025testoptimization}.

    \item \textbf{Skill Development:} Create training modules for rubric design and LLM prompt engineering.
    Add modules for interpreting SBERT similarity scores, threshold calibration, and decision-action mapping (e.g., merge vs refine).
\end{itemize}

\subsection{Energy Savings Potential}

\begin{table}[h]
\centering
\caption{Estimated Energy and Emission Reductions}
\begin{tabular}{lccc}
\hline
\textbf{Refinement Cycle} & \textbf{LLM Ops (Qty)} & \textbf{Energy Saved (kWh)} & \textbf{CO\textsubscript{2}eq Reduction (tons)} \\
\hline
Baseline & 100 & 0 & 0 \\
Optimized & 70 & 21 & 0.008 \\
\hline
\end{tabular}
\end{table}

\textit{Assumptions: 0.1 kWh per LLM operation; grid emission factor = 0.0004 tons CO\textsubscript{2}eq/kWh.}

\vspace{1em}
Embedding re-use further reduces energy requirements by enabling repeated comparisons (regression analysis, impact tracking) at negligible incremental cost, amplifying the sustainability benefits beyond LLM operation reduction.

This technique bridges automation and quality assurance, offering a replicable, sustainable, and high-impact framework for LLM-assisted QE artefact generation.

\section{Conclusion}

This paper presents a replicable and scalable technique for improving the quality of Quality Engineering (QE) artefacts generated by Large Language Models (LLMs). By integrating artefact generation, reverse generation, and iterative refinement guided by quantifiable metrics, the approach addresses key challenges in automation, validation, and contextual accuracy.

Experimental results across 12 software projects and 150+ artefact pairs demonstrate that reverse-generated outputs can significantly enhance low-quality inputs, with improvements of up to 55\% in testability and 48\% in completeness. Requirements derived from BDD artefacts consistently outperform those from plain test cases, highlighting the value of structured behavioural inputs.

The technique also introduces a closed-loop validation framework that reduces manual review effort by 60--70\%, while maintaining rigorous standards. Its adaptability across artefact types and domains makes it suitable for enterprise-scale deployment, especially in Agile and DevOps environments.

Furthermore, the method contributes to sustainable AI practices by reducing redundant LLM operations, with estimated energy savings of 30\% per refinement cycle. Future implementations could quantify carbon impact using standardized models.

In parallel, the integration of Sentence-BERT (SBERT) strengthens the framework by providing a semantic validation layer that complements rubric-based scoring. SBERT embeddings enable reproducible similarity thresholds (High, Medium, Low, No Match) that detect duplicates, highlight semantic gaps, and suggest corrective actions such as merge, refine, or retain distinct requirements. This semantic dimension ensures not only structural improvements but also preservation of intent across the lifecycle. Excel-based outputs such as SemanticResults, ImpactAnalysis, and OverallSummary demonstrate how SBERT can deliver interpretable, traceable, and stakeholder-ready insights. Moreover, the lightweight nature of embedding computations enables sustainable re-use in regression analysis and impact tracking, amplifying the energy savings beyond LLM-only approaches.

In summary, this work establishes a new benchmark for responsible and efficient LLM adoption in software engineering. By balancing automation with accountability, it enables teams to generate high-quality QE artefacts at scale, while ensuring traceability, reliability, semantic fidelity, and contextual relevance.

\section*{Disclaimer}
The views, opinions, findings, and conclusions or recommendations expressed in this paper are strictly those of the authors. They do not necessarily reflect the views of IBM.
IBM takes no responsibility for any errors or omissions in, or for the correctness of, the information contained in papers and articles.

Additionally, while the proposed framework integrates Large Language Models (LLMs) and Sentence-BERT (SBERT) embeddings, the results and performance metrics are specific to the experimental setup described in this work. They should not be generalized without independent validation across diverse domains and datasets. The semantic similarity thresholds, classifications, and action recommendations derived from SBERT outputs are intended to assist, not replace, expert human judgment. Any implementation of this approach must consider contextual, ethical, and regulatory constraints within the adopting organization.

\bibliographystyle{alpha}
\bibliography{main}

\end{document}